# Deep Learning Architecture for Motor Imaged Words

Vimal. W[1] and Akshansh Gupta[2]

[1]Anna University, Chennai, India
`1*vimalwilliam99@gmail.com`

[2]Central Electronics Engineering Research Institute
`2akshanshgupta@ceeri.res.in`

**Abstract.** The notion of a Brain-Computer Interface system is the acquisition of signals from the brain, processing them, and translating them into commands. The study concentrated on a specific sort of brain signal known as Motor Imagery EEG signals, which are activated in the brain without any external stimulus of the needed motor activities in relation to the signal. The signals are further processed using complicated signal processing methods such as wavelet-based denoising and Independent Component Analysis (ICA) based dimensionality reduction approach. To extract the characteristics from the processed data, both signal processing includes Short-Term Fourier Transforms (STFT) and a probabilistic approach such as Gramian Angular field Theory are used. Furthermore, the gathered feature signals are analyzed and converted into noteworthy commands by Deep Learning algorithms, which can be a mix of complicated Deep Learning algorithm families such as CNN and RNN. The Weights of trained model with the particular subject is further used for the multiple subject which shows in the elevation of accuracy rate in translating the Motor Imagery EEG signals into the relevant motor actions.

## 1   Introduction

The Brain-Computer Interface is one field of study that has the potential to push the limits of what is possible for human thought. The processing and translation of brain signals can be done in several different ways. Meng [1] described an approach that combines several different signal processing methodologies to process the incoming brain signals. He also demonstrated weight-based voting of stacked classifiers to analyze and convert the signals into the relevant commands. One may get an idea about the 3D Convolutional Neural Network from Miao [2], which is based on the Spectral-Frequency Temporal method and can analyze motor imagery data in the SFT domain. Miao also discusses other aspects of the network. Li [3] shows how parallel connections between the layers of the neural network may be used to extract spatial and temporal characteristics from motor imaging data. This can increase the accuracy of the signal's interpretation and translation. Zhang [4] proposed an approach of a Hybrid Deep Learning model with the combination of Transfer Learning techniques to ensure that the system works for the various subjects with a limited amount of data. This is made possible by fine-tuning the fully connected (FC) layers in the Deep Learning architecture. Additionally, the proposal states that with the combination of Convolutional Neural Networks with the Long-Short Term Memory to analyze the temporal and spatial features of Motor Imagery signals, the system is able to analyze the temporal and spatial features of Motor Imgery Signals. Cherloo [5] provides an explanation of the Ensembled Regularized Common Spatio-Spectural Patten model, which is superior to the standard Common Spatio-Spectural model, which just has the spatial filters, in terms of its accuracy with regard to the Motor Imagery Signals. According to Yu [6], an enhanced processing approach for the motor and mental imagery signals is described. This processing approach includes Non-Stationary Signal Decomposition, Enhanced Empirical Fourier Decomposition, and Multiscale Principal Component Analysis. The processed signal is then further analyzed using the Feedforward Neural Network model. Tang [7], During the course of his investigation, Tang came up with the hypothesis that the spatial information contained within the signal may facilitate a more accurate translation of the motor imagery signal into an appropriate command. In addition to this, he devised an upper triangle filter for the purpose of further identifying discriminative frequencies and extracting common spatial characteristics. These two activities were carried out as a part of his inquiry into the matter. After the feature has been retrieved, the autoencoder method is applied as the next step in the analysis process.

This study focuses on reducing the mathematical complexity of the processing of the motor imagery signals, which assists in improving the latency of processing and clearing the raw incoming signals from the brain. It is based on the research publications that have been listed above. According to the findings of a number of studies, a significant amount of temporal and spatial feature information derived from motor imagery signals is required for accurate translation of those signals into their corresponding instructions. In order to learn the key temporal and spatial properties of the motor imagery signals, two distinct deep learning algorithms are utilised. These algorithms are fed the features that were extracted from the motor imagery signals. This research also centred on the procedures of noise reduction and frequency domain feature extraction, as well as the analysis of feature extracted signals in one dimensional and two dimensional signal forms. The method of feature extraction is changed between the frequency domain analysis and the probabilistic model for the purpose of converting the signal into a two-dimensional matrix. This is done for each dimension individually

Regardless of the denosing and artifacts removal processes from the raw motor imgaery signals from the brain. The performance of each dimension for the various subjects, as well as the training weights of one subject, are compared with the weights used by the other subjects, and the results are estimated and analysed.

## 2  Proposed methodology

The research into connecting the brain with a computer has placed a strong emphasis on enhancing performance while also reducing the amount of delay experienced in the signal processing. The study works on the various stages of the motor imagery signal analysis with the appropriate mathematical support in order to remove noise and artefacts from the signals. This is necessary because the data acquisition is based on the non-invasion method, which may carriers more noise due to fresh and skull. Consequently, the study works on the various stages of the motor imagery signal analysis. The approach incorporates a number of distinct phases for the treatment of EEG signals, such as signal processing and an examination of the signal's frequency domain.

### 2.1  Data Processing

The Motor Imagery Signals can be triggered without any external motor activities, which activates the neural connection between the neurons in a particular location. This occurs because the Motor Imagery Signals are stimulated. The data on the communication that takes place between the neurons is gathered in the form of motor imagery signals, which are a subtype of EEG signals and are gathered by a procedure that does not involve invasion. The procedure generates a significant amount of noise in the signals as well as the existence of irrelevant motor activities as a result of some disruption. This noise and these unnecessary motor actions need to be handled before the feature extraction techniques can be used. In order to get rid of the undesired noise and artefacts that were present in the MI-EEG data, the research concentrated on signal treatment techniques such as wavelet-based denoising and independent component analysis (ICA). Independent Component Analysis, also known as ICA for short, is an algorithm that is highly recommended for use in the field of neuroscience. ICA is able to transform the MI-EEG signal channels into the most maximally independent variable possible. This is accomplished by combining the statistical independence with the non-gaussian independence, which are both important factors. When applied to a dataset, the ICA allows for the reduction of major artefacts and the identification of truly independent variables.

$$p(x, y) = p(x)p(y) \qquad (1)$$

Where, $p(x)$ respresents the probability distribution of x and $p(x,y)$ respresents the joint probability distribution of x and y. Eq. 1 mathematically describes the statistical independence employed in the ICA. After a significant amount of artefacts have been removed from the M-EEG signals, the signals need to be cleaned of noise, which may be accomplished through the use of a wavelet-based denoising approach. The approach includes a number of processes, such as decomposing the signal and extracting its coefficients. After that, the coefficients are cleaned up using a threshold-based method, and ultimately, the noise-free coefficients are rebuilt into the original signal.

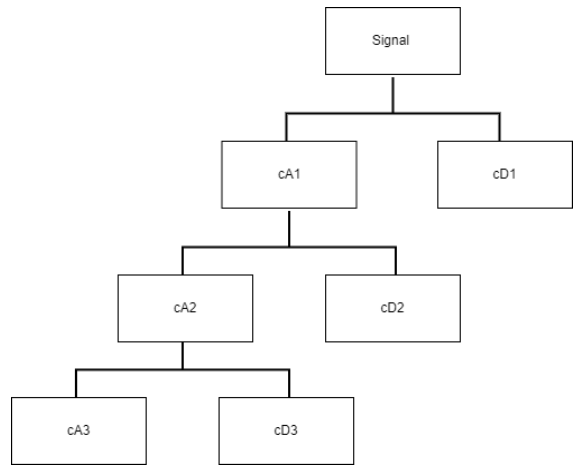

**Fig.1.** Wavelet Decomposition of Signal

Figure 1 provides a description of the 3-level wavelet decomposition that, when applied to the signal, yields the three levels of the signal's coefficients. This decomposition is shown in the figure. The two coefficients that reflect the signal's coefficients of detail and approximation are denoted by the letters cD and cA, respectively. After the decomposition, the updated coefficients are employed for the reconstruction of the signal, and the lowest coefficients related with the noise are essentially ignored.

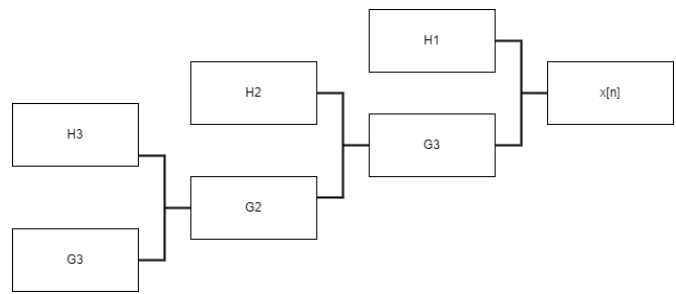

**Fig.2.** Wavelet Reconstruction of Signal

Figure 2 provides a description of the three-level wavelet reconstruction that was applied to the threshold-based coefficients that were deleted because they were related with noise. In addition, the revised set of coefficients that was applied when attempting to rebuild the initial signal. The use of ICA in conjunction with a wavelet-based denoising approach has the potential to lessen the amount of substantial noise and artefacts that are present in the signal, which, in turn, has the potential to increase the accuracy with which the model performs in relation to the data.

## 2.2 Feature Engineering

As a result of the fact that MI-EEG signals may be analysed in both one and two dimensions, the research makes use of a variety of feature engineering approaches. The one-dimensional signals are capable of transforming from the time domain into the frequency domain. This is because frequency domain analysis may be simple to process, and as a consequence, the findings can be improved. In addition, the signals may be turned into a two-dimensional matrix, which is something that can be accomplished by adopting an improved probabilistic model.

**Frequency Domain Anlysis.** The signal may be analysed in the Fourier domain, which can offer a more extensive method of analysing the signal than can be done in the time domain. The full continuous time signal can be translated into the Fourier domain by using short-term Fourier transforms. Consider a signal MI-EEG signal that is continuous in time and it is denoted by x(t). The STFT is computed in such a way that the x(t) is divided into smaller segments and for each segment the Fourier transforms are applied, which returns the frequency and phrase content for each segment over the time.

$$\{x(t)\}(\tau,\omega) \equiv X(\tau,\omega) = \int_{-\infty}^{\infty} x(t)\omega(t-\tau)e^{-i\omega t}dt \qquad (2)$$

Where,
$x(t)$ - Represents the MI-EEG signal to be transformed into fourier domain,
$X(\tau,\omega)$ - Denotes the Fourier transforms over the continuous time signal
$\omega(\tau)$ - Window function

Eq. 2. Mathematically describes the Short-Term Fourier transforms which can be applied over the continous time motor imagery signals. x(t) is transformed into sinusoidal frequencies and the phrase content, this can represented as a spectrogram. Fig.3. graphically describes the STFT spectrogram of the single motor imagery signal for the entire dataframe. Simillary, the entire signals can be converted fourier domain with the application of STFT algorithm over the dataframe which inturns enhance the features present in the signal.

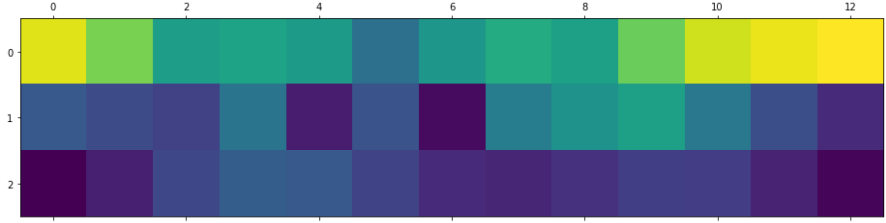

**Fig.3.** STFT Spectrogram

**Probabilistic Model.** The continuous time motor imaging signals may be processed on a two-dimensional platform, which can result in improved performance when it comes to converting motor imagery signal x(t) into applicable commands. The research uses a technique known as Gramian Angular Field (GAF), which plots the time series data in polar coordinates rather than Cartesian coordinates. This allows the researchers to translate the MI-EEG signals to visuals. Identifying the temporal correlations requires either taking into account the trigonometric difference between each point in the data frame or taking the total of those differences.

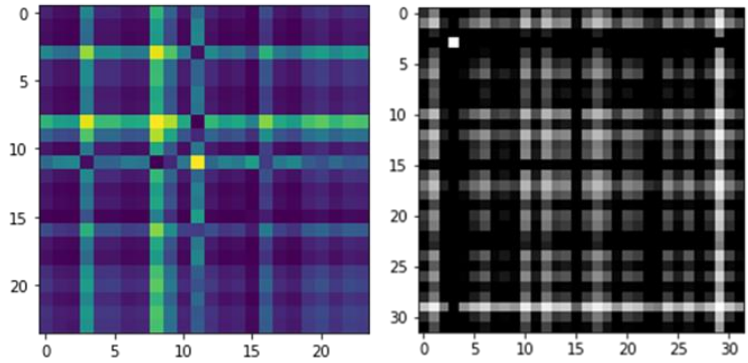

**Fig.4.** Resultant Image of Gramian Angular Field (GAF)

Fig.4 depicts the picture that is produced as a consequence of applying the gramian angular field, which transforms the signal x(t) into the polar 2-dimensional matrix image. This image is the result of applying the gramian angular field.

$$\begin{pmatrix} \cos(\phi_1 + \phi_1) & \cdots & \cos(\phi_1 + \phi_n) \\ \vdots & \ddots & \vdots \\ \cos(\phi_n + \phi_1) & \cdots & \cos(\phi_n + \phi_n) \end{pmatrix} \quad (3)$$

The Gramain Angular Field (GAF) over the motor imagery signal x(t) is mathematically described in Equation 3, and it can be converted into two-dimensional images. The Feature Engineering technique that was used in this study helped to improve the quality of the motor imaging signal, which can now be analysed on a 1-dimensional as well as a 2-dimensional platform. The short-term Fourier transform is a tool that may be used to aid in the analysis of time series signals in the Fourier domain. The GAF provides an expanded version of the analysis of the signal in the form of two dimensions.

**2.3. Artifical Intelligence Framework**

The primary emphasis of the study was an investigation of the three-dimensional structure of the motor imagery signals. The artificial intelligence framework may be developed in such a manner that it can analyse both the 1D and 2D kinds of motor imaging signals after the feature engineering is completed on the signal for both 1D and 2D. It is necessary to have a combined framework architecture in order to take into account the information from the temporal and spatial characteristics that are already present in the feature extraction data. In order to do this, the convolutional layers must be fused with the Long-Short Team Memory (LSTM).

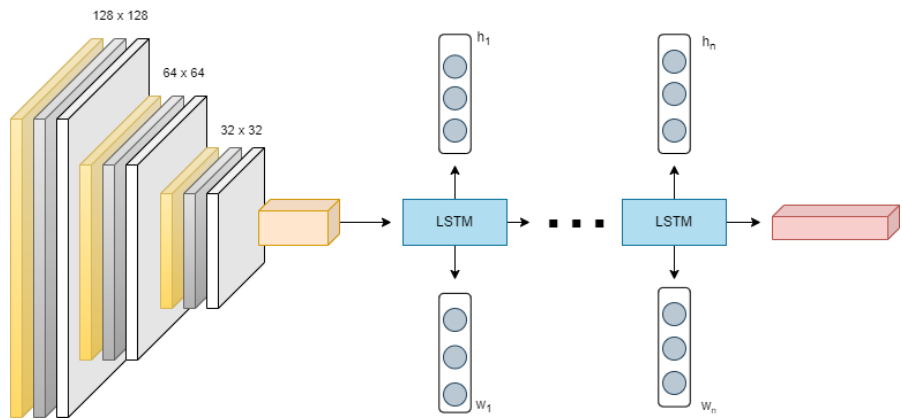

**Fig.5.** Architecture of CNN2D – LSTM

Converting the motor imagery signal x(t) into a two-dimensional image matrix that can then be analysed by combining the convolutional neural network (CNN) and long-short term memory (LSTM) models is an example of a hybrid deep learning model. When the CNN2D-LSTM algorithm is used to the Imaged Signal Matrix, the information contained within the image's temporal characteristics may be analysed by the CNN2D, while the information contained within the image's spatial features can be analysed by the LSTM. Each of the convolutional layers is connected to a specific layer, such as the dropout layer, the Max Pooling layer, or the Activation Layer. Having a Dropout of around 0.5 on each layer helps tremendously to decrease the overfitting of the 2D motor imagery signals that occurs during the analysis of the model. The flatten layer is connected to the Time-Distributed layer, and in addition, the layer is linked to the LSTM blocks that are used for the study of spatial data. In addition, the SoftMax activation function is placed after the fully linked dense layer neural network that comes after the LSTM in the connection chain. Fig. 5. Describes in graphical form the architecture of the CNN2D-LSTM model, which was discussed in the previous section.

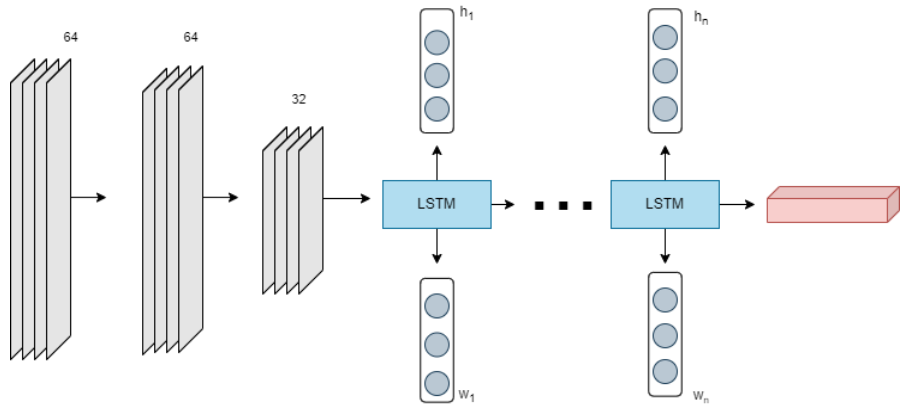

**Fig.6.** CNN1D-LSTM Architecture

The processed motor imaging signal x(t), which is feature extracted using the methodology of the Short-Team Fourier Transform, in which the signal is separated into equal segments and then applied with Fourier transforms to convert it into the Fourier domain. Feature extraction. In addition, both the frequency and the substance of the sentence are represented graphically in a spectrogram, which may be found depicted in figure 3. The analysis of the spectrogram is performed by employing a hybrid model that combines the convolutional neural network (CNN) and the Long-Short Team Memory (LSTM). Specifically, the 1D CNN is fused with the LSTM, and then the dense layers come next. Every CNN layer that is coupled with the activation layer as well as the dropout and max pooling layers has a dropout of 0.2 fixed with it. This is done so that the model does not overfit itself to the data. Figure 6 provides an explanation of the architecture of the CNN1D with the LSTM fused in. An expanded method of analysing the motor imagery signals in multiple dimensions may be obtained by taking into consideration the designs of CNN1D – LSTM and CNN2D – LSTM. The primary goals of this research are to speed up the translation of signals into their associated remarks while also decreasing the amount of time that elapses between each step.

The multiple-dimensional study can provide a comparative examination of the system's performance and latency if it is carried out.

## 3 Results and Discussion

This study's primary objectives were to decrease latency and enhance performance, both of which may be accomplished through comparative research on relevant elements including feature engineering and learning algorithms. Both a one-dimensional and a two-dimensional framework are utilised in the analysis of the motor imagery data. Short-Team Fourier Transforms and Gramian Angular Field Theory are the two main feature engineering methodologies that this study worked with in order to complete each dimensional analysis. Each framework has its own method of feature engineering that must be utilised. Taking into account the significance of neuroscientific data, the learning algorithm was developed in such a way as to extract both the temporal and spatial characteristics from the data.

### 3.1. Artificial Intelligence Framework Analysis

**Convolutional 1D-LSTM.** Combining the Convolutional 1D Neural Network (CNN1D) with the Long-short Team Memory (LSTM) allows for the information to be extracted from the temporal and spatial characteristics of the motor imagery signals. Each Convolutional Layer has an Activation function that is connected with it, such as the ReLu layer, the Max Pooling layer, and the dropout layer. The reduction of models that were overfitting towards the data was made easier thanks to a dropout of 0.2 on each of the convolutional layers. The performance of the model was significantly enhanced by the introduction of new parameters such as the loss function and optimizer. The categorical cross entropy was chosen to represent the loss function, and the RMS prop was chosen to represent the optimizer, both with a learning rate of 0.001. The model is being trained with 70 percent of the data that pertains to motor imagery for a total of 300 iterations. The model is trained and evaluated with the motor imagery data of three separate people, and during these processes, the accuracy, model's latency, and performance are the primary areas of attention, and changes are spoken about.

The accuracies of the three distinct subjects are shown in Figure 7. This figure demonstrates that a progressive increase in accuracy may be accomplished by training the other two subjects with the same weights as were reached after the training of subject 1. The testing of the individual using the thirty percent of unseen data takes around one and a half seconds, which is a significant improvement over the testing of the other two subjects. Therefore, both the reduction in latency and the improvement in performance of the CNN1D-LSTM within the participants have been demonstrated. Table.1. explains performance analysis of CNN1D – LSTM over the motor imagery signals

**Table.1.** Analysis of CNN1D -LSTM model's Performance

| Subject | Class | Precision | Recall | F1-Score |
|---------|-------|-----------|--------|----------|
| Subject 1 | Class 1 | 0.92 | 0.90 | 0.91 |
|  | Class 2 | 0.93 | 0.95 | 0.94 |
|  | Class 3 | 0.93 | 0.94 | 0.94 |
| Subject 2 | Class 1 | 0.95 | 0.93 | 0.94 |
|  | Class 2 | 0.96 | 0.94 | 0.95 |
|  | Class 3 | 0.92 | 0.96 | 0.95 |
| Subject 3 | Class 1 | 0.97 | 0.99 | 0.98 |
|  | Class 2 | 0.98 | 0.97 | 0.97 |
|  | Class 3 | 0.97 | 0.97 | 0.97 |

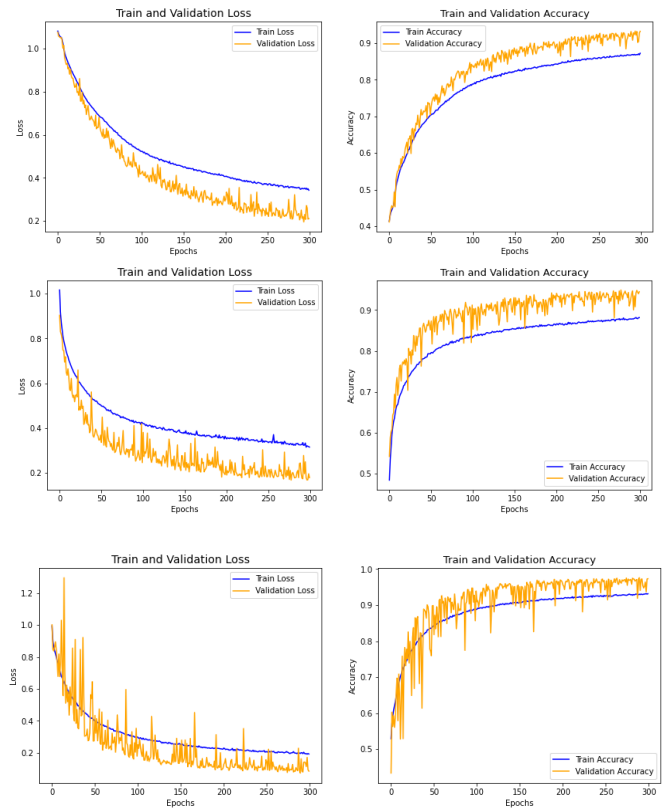

**Fig.7.** Accuracy Vs Epoch Graphs for Different Subjects

During the testing procedure, the Hybrid Deep Learning technique of CNN1D-LSTM over the motor imagery signals demonstrates a greater true positive rate of translating the MI-EEG signals to their appropriate instructions. This is due to the fact that it is applied over the motor imagery signals. Over the course of each topic, the ROC Curve for each class averages approximately 0.99, which represents a progressive improvement in the number of true positives relative to the number of false positives and negatives.

**Convolutional2D-LSTM.** The Gramian Angular Field (GAF) is used to turn the Motor Imagery signals x(t) into a picture. The 2D Convolutional Neural Network is then used to analyse the image, and the layers compress the pixels in order to improve the temporal feature analysis of the picturized Motor Imagery signals. Further, the convolutional layers are fused with the LSTM cells, which assists in the spatial feature analysis of the MI-EEG signals. Additionally, there is a max pooling layer and a dropout layer for each convolutional layer. The dropout value of 0.5 is maintained throughout the convolutional layers, which contributes to a reduction in the amount of signal that is overfit. To achieve more accurate feature analysis, the activation function, also known as ReLu, is coupled to each convolutional layer in the network. RMS prop with a learning rate of 0.001 is used as the optimizer, and categorical cross entropy is used as the loss function in the architectural development of CNN2D -LSTM. The parameters such as the loss function and the optimizer help a great deal in achieving a better translation of signals into commands. In terms of performance and latency, the model obtained an accuracy of around 97 to 99 percent, however in terms of latency, the model requires approximately 5 seconds for the test predictions. The model was trained and tested with the three distinct individuals.

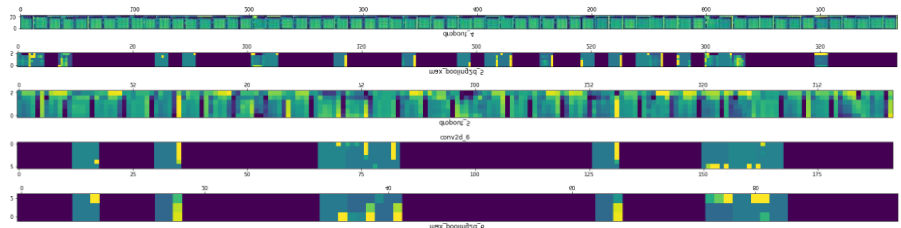

**Fig.8.** Convolutional 2D Feature Map

The feature map of the convolutional 2D is depicted in Figure 8. This is the stage during which the incoming picturized motor imagery signals are analysed by the convolutional layers, which primarily concentrate on the temporal aspects of the data. The diagram illustrates which layers provide an output following a process of convolution, maximum pooling, and application of activation function across a picture of the signal. In comparison to the CNN1D – LSTM, the Convolutional 2D – LSTM has a true positive rate that is marginally higher across the board for all of the study patients. The results of an investigation of the CNN2D-LSTM model's performance are presented in Table.2.

Table.2. Analysis of CNN2D-LSTM model's performance

| Subject | Class | Precision | Recall | F1-Score |
|---|---|---|---|---|
| | Class 1 | 0.99 | 0.98 | 0.99 |
| Subject 1 | Class 2 | 0.99 | 0.99 | 0.99 |
| | Class 3 | 0.99 | 0.99 | 0.99 |
| | Class 1 | 0.99 | 0.99 | 0.99 |
| Subject 2 | Class 2 | 0.99 | 0.99 | 0.99 |
| | Class 3 | 0.99 | 0.99 | 0.99 |
| | Class 1 | 0.97 | 0.99 | 0.99 |
| Subject 3 | Class 2 | 0.98 | 0.97 | 0.97 |
| | Class 3 | 0.98 | 0.99 | 0.99 |

## 4. Conclusion

On Convolutional 1D – LSTM achieves less computational time of 1.5 seconds than Convolutional 2D – LSTM which have about 3 to 5 seconds and in the aspect of accuracy CNN1D – LSTM is about 95 to 97 percent and the CNN2D – LSTM is about 97 to 99 percent on comparing this the accuracies are more or less the same in order. Comparing the parameters such as performance and latency with both dimensional analysis over the motor imagery signals Based on the research conducted regarding Data Processing, Feature Engineering methods, and the Learning Algorithm, it is possible to draw the conclusion that the Convolutional 1D – LSTM algorithm is superior to the Convolutional 2D – LSTM algorithm in terms of both performance and reduction in latency. Based on the findings of the research, it is advised that feature engineering be used in conjunction with the Convolutional 1D-LSTM analysis. Additionally, the pre-processing that is typical of EEG signal processing is strongly suggested. In the future, the study might be directed on improving the 2D analysis of the motor imagery signals and upgrading the feature engineering in order to advance the latency and performance of the algorithm in relation to Motor Imagery Signals.